# AI Regulation in the European Union:
# Examining Non-State Actor Preferences

**Jonas Tallberg, Stockholm University**

**Magnus Lundgren, University of Gothenburg**

**Johannes Geith, Stockholm University**

**Word count: 10,578**

**Abstract:** As the development and use of artificial intelligence (AI) continues to grow, policymakers are increasingly grappling with the question of how to regulate this technology. The most far-reaching international initiative is the European Union (EU) AI Act, which aims to establish the first comprehensive, binding framework for regulating AI. In this article, we offer the first systematic analysis of non-state actor preferences toward international regulation of AI, focusing on the case of the EU AI Act. Theoretically, we develop an argument about the regulatory preferences of business actors and other non-state actors under varying conditions of AI sector competitiveness. Empirically, we test these expectations using data from public consultations on European AI regulation. Our findings are threefold. First, all types of non-state actors express concerns about AI and support regulation in some form. Second, there are nonetheless significant differences across actor types, with business actors being less concerned about the downsides of AI and more in favor of lax regulation than other non-state actors. Third, these differences are more pronounced in countries with stronger commercial AI sectors. Our findings shed new light on non-state actor preferences toward AI regulation and point to challenges for policymakers balancing competing interests in society.

As the development and use of artificial intelligence (AI) continues to grow, policymakers are increasingly grappling with the question of how to regulate this technology. While national authorities were the first to initiate regulation of AI, recent years have seen the emergence of a variety of regulatory initiatives at regional and global levels. [1] This shift reflects a growing realization that AI development often is carried out by companies with transnational activities and creates externalities that do not follow national borders, calling for an international regulatory response.

The most far-reaching international effort to regulate the development and use of AI technology is the European Union (EU) AI Act, proposed by the European Commission in 2021 and currently at the concluding stage of negotiation between the Council and the European Parliament. [2] The EU AI Act will introduce a common European regulatory framework encompassing all sectors and all types of AI technology, except military systems. It will be binding in its nature and regulate the development and use of AI by establishing differentiated rules based on the level of risk involved. While the scope of the Act in the first instance is limited to the EU, there is an expectation that the law might become standard setting globally, much like the General Data Protection Regulation (GDPR).

[1] Council of Europe (2022).
[2] European Commission (2021).

While the EU AI Act formally is negotiated among the EU's institutions, the importance of the Act for future AI development has mobilized large numbers of non-state actors, seeking to influence the terms and conditions of the new regulatory framework. For business actors involved in AI development, the Act will have significant implications for their innovation potentials and competitive positions. For other types of non-state actors, such as non-governmental organizations (NGOs), research institutes, and labor unions, the Act raises critical questions about the protection of individual rights and public interests.

Identifying what these actors want and why is imperative. Non-state actors tend to exert significant influence in EU decision-making[3] and global policymaking generally [4] through lobbying of domestic governments and international institutions. Such influence is likely to be particularly pronounced on an issue such as AI regulation, which presents tech companies with informational advantages, other businesses with strong commercial interests, labor unions with existential fears of automation, and NGOs with opportunities to mobilize broad public concerns. Establishing the preferences of non-state actors is therefore crucial for researchers interested in the drivers of AI regulation, but also for policymakers tasked with balancing competing concerns in society.

---

[3] Klüver (2013); Dür et al. (2019).
[4] Fuchs (2007); Tallberg et al. (2018).

Yet, so far, we know little about the preferences of non-state actors toward AI regulation. Whereas a number of studies have examined state preferences[5] and citizen preferences[6] toward AI regulation, our knowledge about the preferences of non-state actors is limited. While tech executives, labor leaders, and NGO representatives are becoming increasingly vocal about AI regulation, systematic research into their preferences is lagging behind. The interests of business actors are particularly obscure. On the one hand, they have consistently called for more regulation of AI compared to the status quo.[7] On the other hand, businesses have raised concerns about the EU AI Act impeding competitiveness,[8] and even actively lobbied the EU to reduce the regulatory burden.[9] The lack of knowledge is compounded by the possibility of differences across business actors, as leading AI developers, such as tech companies, may harbor regulatory preferences that differ from those of other business actors.

The purpose of this article is to offer the first systematic analysis of non-state actor preferences toward international regulation of AI, focusing on the case of the EU AI Act. What are the core concerns and regulatory preferences of non-

---

state actors with respect to European AI regulation? Why do actors differ with regard to these concerns and preferences?

Theoretically, we develop an argument about the preferences of non-state actors toward AI regulation. We distinguish analytically between two types of non-state actors: business actors, which are driven first and foremost by for-profit motives, and other types of non-state actors, which are driven primarily by non-profit motives. Identifying innovation versus protection as the core dimension of conflict on AI regulation, we argue that business actors and other non-state actors are likely to hold systematically different preferences. Compared to other non-state actors, business actors are less likely to express concerns about AI development and more likely to favor innovation over protection. In addition, we theorize that these differences between business actors and other non-state actors are conditioned by the level of AI uptake in a country, specifically, the strength of the commercial AI sector.

Empirically, we test these expectations using data on non-state actor preferences drawn from the public consultations on European AI regulation organized by the European Commission in 2020, one year prior to the tabling of the EU AI Act. Public consultations offer unique opportunities to identity the regulatory preferences of non-state actors.[10] In all, we analyze a sample of 505 submissions by businesses, NGOs, research institutes, and other non-state actors located in the EU. We examine our hypotheses using descriptive and regression analyses of the expressed concerns and regulatory preferences of these non-state actors.

---

[10] McKay and Yackee (2007); Bunea (2013).

Our core findings are threefold. First, all types of non-state actors express concerns about the implications of AI and support EU regulation involving a variety of mandatory requirements. Second, there are nevertheless significant differences across types of non-state actors, where business actors are less concerned about the downsides of AI and more in favor of lax regulation than other types of non-state actors. Third, the differences between business actors and other non-state actors are more pronounced in countries with stronger commercial AI sectors than in countries with lesser developed AI sectors. In all, these findings suggest that non-state actors generally recognize the need for a common European regulatory framework, but attach systematically varying importance to innovation versus protection depending on actor motives (group type) and competitive position (country).

Our findings have several broader implications. To start with, they contribute to the literature on actor preferences toward AI regulation, providing new evidence on key patterns in non-state actor interests, thus complementing previous research into state and citizen interests. In addition, they suggest that the growing field of research on international AI governance, which so far has focused mainly on states and institutions, would benefit from greater attention to the non-state actors that work to influence these arrangements. Finally, our results highlight the types of political challenges that policymakers confront when developing AI regulation, having to reconcile the competing interests of non-state actors whose support likely is critical for effective and legitimate AI governance.

## Theorizing Non-State Preferences Toward AI Regulation

What are core concerns that inform and shape actor preferences toward AI regulation? In this section, we begin by reviewing existing research on actor preferences toward AI regulation, concluding that the preferences of non-state actors so far have received limited systematic attention. We then turn to our own theoretical argument, centered on the preferences of business actors and other non-state actors toward innovation and protection through AI regulation.

### The State of the Art

Recent years have seen the emergence of a growing body of research on AI governance,[11] albeit to a lesser degree with a focus on the international level.[12] A prominent theme in this literature has been the multiple options for how to govern AI that have been proposed by governments, businesses, NGOs, and academics, and which broadly range from soft-law standards to hard-law regulations.[13] Increasingly, research has also turned to the issue of actor preferences toward AI regulation. Simplifying slightly, we may divide this literature into three parts, depending on the actor in focus: states, citizens, or non-state actors.

---

[11] For overviews, see Bullock et al. (2022); Büthe et al. (2022).
[12] Tallberg et al. (2023).

Since states are in a key position to shape AI governance, studies have sought to map and explain governments' diverse approaches to AI regulation at domestic and international levels, using both in-depth case studies and comparative analyses. Research suggests that governments differ significantly in how they interpret their role and responsibility in AI governance.[14] Governments can be broadly distinguished in either taking a proactive or passive stance towards the development of AI while at the same time either focusing on the regulation of AI risks or the promotion of its deployment.[15] So far, most studies have focused narrowly on the main AI powers, the US and China.[16] Because of the EU AI Act, European AI regulation, too, is beginning to attract considerable attention.[17] Several studies compare the EU's AI approach to that of the US and China,[18] as well as the UK.[19]

---

[13] Cihon et al (2020); Büthe et al. (2022); Schmitt (2022); Stix (2021).
[14] For an overview, see Radu (2021).
[15] Djeffal et al. (2022).
[16] Allen (2019); Ding (2018); Rasser et al. (2019); Roberts et al. (2021b); Hine and Floridi (2022).
[17] For an overview, see Ulnicane (2022).
[18] Roberts et al. (2021a); Roberts et al. (2023).
[19] Cath et al. (2018).

Next to state preferences, citizen attitudes toward AI are gaining growing attention. As AI is implemented in various areas of everyday life, citizens are becoming increasingly aware of its positive and negative consequences. Poorly functioning AI systems in some countries, such as Australia and the Netherlands, and the release of ChatGPT in November 2022, with its consequences for education and content production, have made AI and its regulation a topic of broad public debate. Researchers have therefore begun to examine citizen attitudes toward AI technology and regulation. [20] For instance, studies have explored citizens' regulatory preferences in the EU,[21] the US,[22] the UK,[23] and Germany.[24] In this vein, König et al. (2023) show that German citizens support moderate to strong measures when specifically asked about two core challenges of AI, namely, transparency and energy efficiency.[25] Ehret (2022) offers a first comparative examination, which focuses on five countries (Chile, China, Germany, India, and the UK) and shows that citizen preferences are shaped both by normative aspects and economic consequences of AI systems.[26]

In contrast, we still have very limited knowledge about the preferences toward AI regulation among non-state actors, despite the instrumental role of businesses in developing AI technology, the interests of labor unions to protect worker interests, and the efforts of NGOs to shape public debate on this topic. In one recent contribution, focused on strategies rather than preferences, Auld et al. (2022) show how corporate and civil society actors use private governance venues to engage and push states to institutionalize rules for ethical AI. Yet what these actors want and why remains an open question, whose answer is bound to affect the nature and stringency of emerging AI regulation.[27]

---

[20] For an overview see Zhang (2022).
[21] European Commission (2017).
[22] Zhang and Dafoe (2019).
[23] Ada Lovelace Institute & The Alan Turing Institute (2023).
[24] Kieslich et al. (2022).
[25] König et al. (2023).
[26] Ehret (2022).

*The Argument*

We present our argument in three steps. First, we identify innovation versus protection as the central dimension of conflict in debates over the regulation of AI. Second, we develop our expectations about the regulatory preferences of business actors and other non-state actors on this dimension of conflict. Third, we explain why we expect the strength of the AI sector in a country to condition the regulatory preferences of non-state actors.

Our argument is anchored in rationalist theories of preference formation, which understand preferences as the way an actor orders possible outcomes on a given issue.[28] Preferences are assumed to be complete (i.e., actors are capable of choosing between two or more outcomes) and transitive (i.e., those choices are internally consistent). Theories arrive at preferences in three principal ways: by assumption, observation, or deduction.[29] Our argument is based on the method of deduction, since we draw on general theories of non-state actor preferences to derive expectations about the likely preferences of business actors and other non-state actors toward AI regulation.

______________________________

[27] Auld et al. (2022).
[28] Arrow (1952); Hansson and Grüne-Yanoff (2022).

In rationalist models, preferences are ordered along one or several dimensions of contestation. Previous analyses suggest that the EU political space, for instance, contains multiple dimensions of political conflict. Examples include left versus right,[30] more versus less integration,[31] fiscal transfer versus fiscal discipline,[32] and progressive versus conservative values.[33]

With respect to AI regulation, we assume that the key dimension of political conflict pertains to the trade-off between innovation and protection. This dimension captures different preferences with respect to how regulation of AI should strike the balance between two objectives often perceived to be in tension: one the one hand, creating a regulatory environment that promotes innovation in AI development, and one the other hand, introducing regulation that protects the safety, rights, and values of citizens.

---

This dimension relates to a classic debate about the relationship between regulation and innovation. In research, scholars discuss whether regulation primarily serves to stifle innovation by introducing burdensome requirements, or whether regulation in fact may facilitate innovation by establishing a level and predictable playing field.[34] In politics, countries have chosen to strike different balances between regulation and innovation; while European policymakers are increasingly willing to regulate risks on precautionary grounds, US policymakers are more reluctant to impose additional regulatory controls on business.[35] It is not unlikely that the relationship between regulation and innovation is more complicated than a simple trade-off. For our purposes, the key issue is the perception among actors that AI regulation involves a tension between promoting innovation and ensuring protection.

We find ample evidence of this perception in debates over the EU AI Act and AI regulation generally. The European Commission's proposal for a regulation speaks of how the EU's approach needs to deal with "the twin objective of promoting the uptake of AI and of addressing the risks associated with certain uses of such technology."[36] Member state negotiations in the Council are centered on the trade-off between technological development and risk protection in an effort to strike "a delicate balance".[37] Debates in the European Parliament revolve around the competing goals of ensuring an innovation-friendly regulatory environment and safeguarding the rights and interests of European citizens against the risks of AI.[38] Many other proposed frameworks of AI governance, while varying in scope, call for a similar balance between innovation and harm mitigation.[39]

---

[34] E.g., Vogel (1997); Blind (2012); Aghion et al. (2021).
[35] Vogel (2012).
[36] European Commission (2021, 1).

The innovation versus protection dimension also characterizes other recent EU legislation on data governance. Examples includes the GDPR, which set a precedent for regulating data, as well as the Digital Markets Act (DMA) and the Digital Services Act (DSA), with the goals of increasing European innovation while enhancing individuals' rights over online content.

The core of our argument pertains to differences in expected preferences between business actors and other non-state actors with respect to the appropriate balance between innovation and protection. Non-state actors are a broad category and encompass all actors that are not funded by, directed by, or affiliated with a government.[40] Both analytically and empirically, non-state actors overlap extensively with interest groups.[41] We distinguish between business actors, on the one hand, and other non-state actors, on the other hand. Previous work that examines private governance initiatives around ethical AI standards relies on a similar dichotomy between business and civil society actors,[42] as does earlier research on interest groups and lobbying in domestic and international politics.[43]

---

[37] Council of the EU (2022, 1).
[38] Euractiv, 15 November 2021, "European Parliament, Countries Want More Innovation, Less Burden in AI Act"; Euractiv, 13 February 2023, "AI Act: All the Open Questions in the European Parliament"
[39] Nitzberg and Zysman (2022).
[40] Josselin and Wallace (2001); Avant et al. (2010); Tallberg et al. (2013).
[41] Beyers (2008); Bloodgood (2011).

Business actors, comprising both individual companies and business associations, are for-profit actors, which we assume are primarily driven by the goal to make money, consistent with the neoclassical theory of the firm (Marshall 1890). Generating profit is the over-riding concern of individual companies, while it is an indirect goal of business associations, tasked with protecting the commercial interests of their corporate members.

Other non-state actors, in contrast, are non-profit actors, which we assume are primarily guided by alternative concerns. NGOs, social movements, and philanthropic foundations are conventionally described as driven by values and principles, even if they are often instrumental in their pursuit of these objectives.[44] Scientific actors, such as research institutes and networks, are primarily engaged in knowledge creation and diffusion.[45] While labor unions similarly to business associations seek to protect the interests of their members, those interests involve concerns that often are in tension with profit maximization, such as decent wages, job security, and good working conditions.[46]

---

[42] Auld et al. (2022).
[43] E.g., Mahoney (2008); Dür et al. (2019); Ibid. (2023).
[44] Sell and Prakash (2004); Mitchell and Schmitz (2014).
[45] Haas (1992); Miller (2007).
[46] Ahlquist (2017).

Building on these assumptions, we expect business actors and other non-state actors two hold different preferences, on average, when approaching the issue of how AI should best be regulated at the European level. When confronted with the choice between innovation and protection, we would expect business actors to be relatively more in favor of innovation than other non-state actors, which, conversely, would be relatively more in favor of protection. This expectation resonates with recent research on non-state actor preferences toward international trade agreements, which points to distinct variation between business actors and other non-state actors in their support for regulatory measures, reflecting differences in core concerns among these actors.[47]

A regulatory environment that favors innovation is likely to be perceived by business actors as more conducive to their commercial interests in AI development and use. On balance, business actors are likely to prefer more permissive rules, lower bureaucratic hurdles, and fewer regulatory restrictions. In such a regulatory environment, European firms will enjoy greater room for maneuver as they seek to develop AI applications at the international forefront, resulting in a stronger position vis-à-vis competitors in China and the US.

This is not to say that business prefers AI development and use to be unregulated. Indeed, AI business leaders, such as OpenAI's Sam Altman, Google's Sundar Pichai, and DeepMind's co-founder Mustafa Suleyman, have all spoken up on the need for AI to be regulated. According to Pichai, for instance, "AI needs to be regulated in a way that balances innovation and potential harms."[48] Yet, when lobbying policymakers, business typically argues in favor of regulatory arrangements involving more voluntarism and less stringency[49]. When the European Parliament adopted its final position on the EU AI Act in June 2023, this was greeted by an open letter from about 150 business executives calling for laxer regulation, or else Europe would "miss the chance to rejoin the technological avant-garde."[50]

---

[47] Dür et al. (2023).

Other non-state actors are likely to be less enthusiastic about a regulatory environment perceived to favor business innovation over public protection. Instead, NGOs, research institutes, and labor unions are more likely to prefer European rules for AI development and use that prevent undue risks, safeguard the public interest, and ensure respect for fundamental rights, including privacy and non-discrimination.

This does not to mean that such non-state actors are insensitive to the commercial importance of AI. Labor unions, for instance, tend to emphasize that AI both offers an economic potential for companies on which workers are dependent *and* presents a challenge to the jobs and rights of workers (ETUI 2023).

---

[48] Financial Times, 23 May 2023, "Google CEO: Building AI Responsibly Is the Only Race that Really Matters"
[49] Time, 20 June 2023, "OpenAI Lobbied the EU to Water Down AI Regulation"
[50] Financial Times, 30 June 2023, "European Companies Sound Alarm Over Draft AI Law"

Likewise, research institutes and think tanks frequently trumpet the potential of AI technology, while underlining the need for safe development and deployment (Stanford Graduate School of Business 2018; Tony Blair Institute for Global Change 2023). On balance, however, non-business actors tend to put a greater emphasis on the risks of AI and the need for protection.

These expectations translate into two hypotheses about anticipated differences between business actors and other non-state actors in their approaches to AI regulation, as captured by the innovation versus protection dimension.

The first hypothesis focuses on the *concerns* expressed with respect to AI technology. By concerns we mean the worries that actors have with regard to possible negative consequences of AI, such as endangering of safety, breaches of fundamental rights, and discriminatory outcomes. We expect that business actors are less concerned with potential downsides of AI development and use than other non-state actors, since business actors privilege the commercial opportunities offered by AI.

*H1*: Business actors are less likely to express concerns about AI technology than other actors.

The second hypothesis extends this logic to the *regulatory preferences* of non-state actors toward AI regulation. By regulatory preferences we mean the expressed interests of actors with respect to the restrictiveness of rules governing the development and use of AI. We expect that business actors are more in favor of laxer regulation of AI technology than other non-state actors, since business actors are anxious to ensure an innovation-friendly regulatory environment.

*H2*: Business actors are more likely to express preferences for laxer regulation of AI than other actors.

We have so far assumed that business actors and other non-state actors operate in identical environments. In practice, however, the uptake of AI varies across countries.[51] Building on basic notions in political economy, we expect such differences to matter for the perspectives of non-state actors on AI development and regulation. Specifically, we anticipate that the strength of the commercial AI sector in a country conditions the concerns and preferences of business actors and other non-state actors in varying but predictable ways.

Business actors in a country with a more developed commercial AI sector are better positioned to benefit from an integrated European AI market than business actors in a country with a less developed sector. Business actors in more developed AI environments have likely benefited from network effects, competitive pressures, and commercial developments that give them an edge when entering a level European playing field. For the same reasons, business actors in less developed AI settings are likely to be worse prepared to compete on an integrated European AI market.

---

[51] Fatima et al. (2021); Tortoise (2023).

Turning to other non-state actors, we can expect a similar pattern, but driven by other dynamics. In countries with stronger commercial AI sectors, other non-state actors are more likely already to have encountered issues related to protection, making them more attuned to the risks of AI development. In comparison, other non-state actors located in countries with weaker commercial AI sectors are less likely to have experienced the potential downsides of AI development.

Combining these dynamics, we would expect the expressed concerns and regulatory preferences of business and other non-state actors to vary based on the *strength of a country's commercial AI sector*. By implication of this logic, the gap between concerns and preferences would widen as we move from less to more developed commercial AI environments.

*H3*. The more developed the commercial AI sector in a country, the greater the differences in expressed concerns between business actors and other non-state actors.

*H4.* The more developed the commercial AI sector in a country, the greater the differences in regulatory preferences between business actors and other non-state actors.

## Data and methods

To test our hypotheses, we identify the regulatory preferences of non-state actors based on responses submitted within the EU public consultation on the White Paper that presented policy and regulatory options for the AI Act. The public consultation was open for submissions between February 20 and June 14, 2020, and the intention was to consult stakeholders with an interest in AI, including AI developers, businesses and business associations, NGOs, public administrations, academic institutions, and private citizens.[52] The EU Commission was especially interested in how respondents viewed the impact of AI on safety and liability regimes, and what regulatory options they preferred. This process was conducted through an online platform where stakeholders could submit their comments and suggestions, both as open-ended answers and closed-form numerical responses to specific questions posed by the EU Commission. We chose to focus on this public consultation for three main reasons. First, throughout the legislative process of drafting and negotiating the EU AI Act, the public consultation on the White Paper was the most comprehensive in scope. Later consultations received fewer submissions. Second, while we recognize that the debate on AI regulation has developed further in response to technological development and the accentuated political salience of AI, the 2020 public consultation allows us to investigate broad stakeholder concerns with regard to AI and assess how these concerns are reflected in general regulatory preferences. Later consultation procedures focused more narrowly on specific legislative proposals. Third, the closed-form numerical responses allowed us to quantify information on a large number of different stakeholders and analyze them comparatively.

---

[52] European Commission (2023).

Using public consultation submissions as a source of data on regulatory preferences is a well-established approach in research on non-state actors in the EU[53] and other national and international contexts.[54] As explained by Bunea (2013), EU public consultations represent a formalized dialogue between policymakers and non-state actors taking place at the policy formulation stage, where lobbying and interest group activity is typically the most intense.[55] For this reason, they constitute a suitable basis for measuring the regulatory preferences of non-state actors.

One possible concern in using EU public consultations as data on regulatory preferences is the risk of bias in stakeholder participation. The EU has different consultation procedures that allow for the involvement of non-state actors that subsequently affect what kind of actors gain access to these procedures,[56] how this relates to the diversity of groups involved[57] and what value stakeholders may have from participating in various consultation formats.[58] While EU institutions seek to ensure that the consultation process is inclusive and transparent, open to a broad range of actors, and do not pose significant resource constraints, the possibility of biased participation cannot be excluded. While some researchers indicate that the Commission has successfully managed to alleviate stakeholder bias,[59] others have found that participation is skewed in favor of business interests and that bias is accentuated in consultations on policy issues that are non-salient and technically complex.[60]

---

[53] Bunea (2013); Klüver (2011).
[54] E.g., McKay and Yackee (2007).
[55] Bunea (2013).

Since the population of relevant stakeholders in the AI policy domain is unknown, we cannot determine the risk of stakeholder bias in our specific sample. The issue of AI is technical, which may increase bias, but has also been a salient topic of popular and policy discussion, which may alleviate bias. The observed distribution of participation does not suggest grave asymmetries across actor type or geographic locations (see below). The possible exception is the large presence of groups based in Belgium, which is to be expected due to the location of the EU headquarters in Brussels, leading many non-state actors to establish a formal presence there as basis for lobbying activities.[61] Overall, while we believe that our sample is reasonably representative of non-state actors that typically contribute to EU public consultations, we are cautious in generalizing our results beyond the participating organizations.

---

[56] Arras and Beyers (2019).
[57] Fraussen (2020).
[58] Binderkrantz et al. (2022).
[59] Quittkat and Kotzian (2011); Bunea (2017); Binderkrantz et al. (2020).
[60] Rasmussen and Carroll (2014); Røed and Hansen (2018).

The public consultation on the EU AI Act received a total of 1,216 contributions. Of these, we exclude 460 responses that lack information on the identity of the stakeholder. Given our theoretical interest, we also exclude responses from non-EU entities (119) and private citizens (132), but we report results where the former category is included in our robustness tests. Our final sample includes 505 responses by entities located within the EU, including non-EU entities that report an office or headquarters within the EU, submitted in a variety of languages.[62] Table 1 shows the distribution of responses in the sample across actor type. We note that about 40 percent of responses were received from stakeholders that we classify in the business category, which includes both business associations and individual firms, while other groups make up the remaining 60 percent. As shown in Table 2, if we focus on the responses submitted by business actors only, slightly more than a quarter (25.9 percent) came from actors operating in the field of technology and innovation ("tech") and the remainder other sectors.

---

[61] Our results are robust to excluding submissions by actors based in Belgium. See Table A.10 in the online appendix.

[62] See Table A.12 in the online appendix for the distribution of responses from EU -countries. A small number of multinational corporations and business associations headquartered outside of the EU that report inside-EU locations in their submissions are included in the sample. For example, Fujitsu, a Japanese firm has reported its location as Belgium, where it has a presence.

**Table 1. Distribution of actors in sample, by actor type**

| Type | Frequency | Proportion (%) |
| --- | --- | --- |
| Academic | 85 | 16.8 |
| Business | 201 | 39.8 |
| NGO | 109 | 21.6 |
| Other | 110 | 21.8 |

Note: N=505. The academic category includes "Academic/Research institutions"; the business category includes "Company/Business organization" and "Business Association"; the NGO category includes "Consumer organization", "NGO (Non-governmental organization)"; the other category includes "Trade Union", "Public authority", and "Other".

**Table 2. Distribution of business actors in sample, by sector**

| Sector | Frequency | Proportion (%) |
| --- | --- | --- |
| Technology and innovation ("tech") | 52 | 25.9 |
| Manufacturing and industrial | 22 | 11.0 |
| Retail and consumer goods | 14 | 7.0 |
| Services and consulting | 61 | 30.4 |
| Other | 52 | 25.9 |

Note: N=201.

We identify policy issues based on the Commission's consultation questionnaire, focusing on the two clusters of questions that pertain to the policy issues for which we have developed theoretical expectations (see Section A.1. in the Appendix). First, we identify the policy issue *concerns about AI*, which encompasses questions relating to whether AI may endanger safety (F25), breach fundamental rights (F26), lead to discriminatory outcomes (F27), take unexplainable actions (F28), complicate compensation for harm (F29), and be inaccurate (F30).[63] Responses to these questions are submitted on a numerical scale, 1-5, with 5 indicating that the respondent considers that a specific concern is "very important" and 1 as "not important at all."

A second policy issue is formulated as *regulatory stringency*, which encompasses questions relating to the preferred design of the regulatory provisions of the AI Act, specifically the importance of mandatory requirements regarding the quality of training datasets (F39), the keeping of records and data (F40), information on the purpose of AI systems (F41), robustness and accuracy of AI systems (F42), human oversight (F43), and clear liability and safety rules (F44). Responses to these questions are analogously recorded on a 1-5 scale.

The policy preferences of each respondent to the questionnaire are indicated by the submitted values (1-5) on these dimensions of concern and regulatory design. For example, a response of "5" on question F43 is assumed to indicate a strong policy preference in favor of the EU AI Act including mandatory requirements for human oversight in AI systems. This approach to measuring policy preferences is consistent with previous research in non-state actor influence (e.g., Bunea 2013) and EU decision-making (e.g., Lundgren et al. 2019).

---

[63] Table A.1 in the appendix provides further detail on the questions.

In our analyses, policy preferences are reflected in two dependent variables, observed at the level of non-state actor consultation submissions. The first variable measures the level of *concerns about AI* and is calculated as the unweighted mean of each respondent's submitted scores on the questions pertaining to AI concerns (F25-30). The resulting interval variable can take values between 1 and 5, where lower values correspond to a lower level of general concern about the risks of AI and higher values indicate a higher general concern. The second dependent variable measures *regulatory stringency* and it is analogously created as the unweighted mean of the responses to questions F39-44, with higher values corresponding to a preference for a more demanding AI regulatory framework and lower values to a preference for a laxer framework. In our robustness checks, we present results where the constituent components (questions) are used as dependent variables.

On the explanatory side, we include a categorical variable to represent *actor type*, which records the type of the observed non-state actor (see Table 1). To facilitate substantive interpretation, we in some models employ a dichotomous variable, *business actor*, which takes the value of 1 if the observed actor is a business association or individual firm, and 0 otherwise. In some models, we also disaggregate business actors into two categories based on their main economic activity, contrasting *tech* actors against those active in other areas (such as manufacturing and industrial; retail and consumer goods; and services and consulting).

We measure the strength of a country's AI sector based on data from the Global AI index, which benchmarks countries on their level of investment, innovation, and implementation of AI technologies.[64] We focus on the commercial component of the index, which reflects the level of AI startup activity and AI investment and business initiatives in the non-state actor's reported headquarter country. The index component comprises 17 individual indicators, including the number of AI companies per capita, funding to AI startups, and number of AI listed companies. Non-state actors from countries with more developed commercial AI sectors will score higher on this variable.

We estimate unconditional and conditional differences between groups using linear regression models with heteroskedasticity-robust errors clustered on country. Due to missingness in responses, our main regression models are estimated on a sample varying between 404 and 419 responses (and where only business actors are concerned, between 156 and 168 responses). In the robustness tests, we report coefficients for alternative specifications and estimators, including models with individual questions, country dummies, multilevel models, country-level controls, and models fitted on samples that exclude organizations based in Belgium or which include non-EU actors (Tables A.5-A.11 in the online appendix).

---

[64] The Global AI index is available via https://www.tortoisemedia.com/intelligence/global-ai/ [last accessed on February 20, 2023]

# Results

We begin our empirical analysis by presenting some descriptive analyses of patterns that emerge when actor responses are aggregated at the country level. Figure 1 shows the mean value of the two key dependent variables for the submitted actor responses, across the headquarter countries in the sample. We make two key observations.

First, on both measures, scores are considerably closer to their maximum (5) than the lower end of the scale (1). This indicates that, on average, non-state actors consider concerns about AI as "important" to "very important," and that they correspondingly consider it "important" to "very important" to include a range of mandatory requirements in the EU AI Act. Across all groups and countries, the mean score is 4.3 for concern about AI and 4.5 for regulatory stringency. In other words, non-state actors who participated in the EU's public consultations must be considered quite worried about the implications of AI and are supportive of relatively demanding regulation.

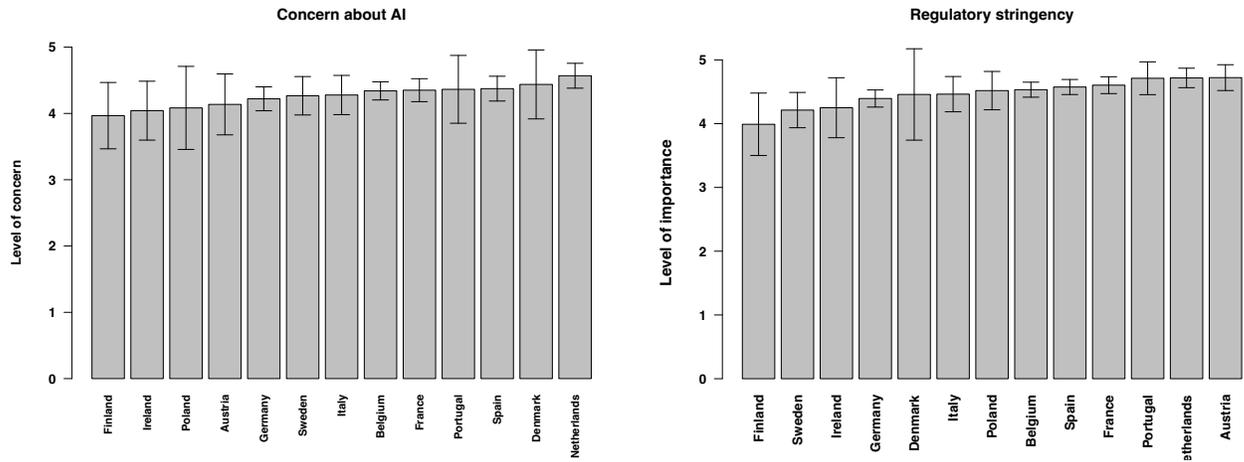

**Figure 1.** Mean level of concern about AI (left) and mean level of preferred regulatory stringency (right) of non-state actor submissions, by country of reported headquarter. Error bars indicate 95 percent confidence intervals. Countries with fewer than five submissions not shown. Data: European Commission 2023.

Second, while differences in mean scores across actors headquartered in different countries are relatively modest, there are interesting patterns of variation and co-variation. It is clear that actors based in some countries, such as Finland, hold views that are considerably more AI-friendly than others, both in terms their views of the risks of the technology and how it should be regulated. In general, actors located in countries with low means on concern tend to have lower values on regulatory preferences, and vice versa, suggesting that the level of concern about AI is correlated with regulatory preferences. This is consistent with the interpretation that regulatory preferences are partly a function of the level of concern about AI.

Turning to our regression analysis, Figures 2 through 6 exhibit the principal results in the form of adjusted predictions.[65] Our first hypothesis was that business actors would exhibit lower levels of concern about AI. As can be seen in Figure 2, the data are consistent with this conjecture. The predicted mean level of concern by business actors is 3.92, which is considerably lower than that of academic institutions (4.39), NGOs (4.60), or other actors (4.57). The differences between business actors and the other groups are statistically significant (p<0.01).

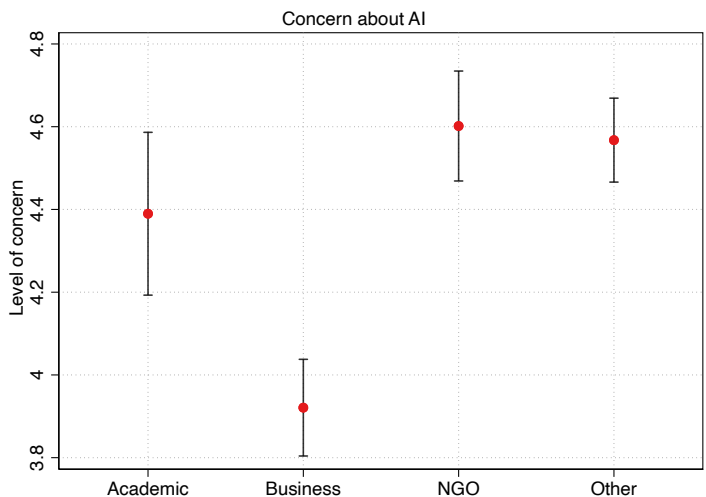

**Figure 2.** Adjusted predictions of group type on level of concern about AI (1-5). Higher values correspond to a higher concern. Average marginal effects with 95

percent confidence intervals. Calculation based on Model 1 in Table A.2. Standard errors clustered on countries. N=411.

The contrast between business actors and other actors is reflected in the content and orientation of the qualitative submissions to the public consultation. A comment submitted by Thales SA, a large French business actor in the aerospace sector, exemplifies this: "As a general remark concerning this EU consultation, the emphasis seems to be put more on concerns than on opportunities. Highlighting examples of beneficial impact and added-value would be appropriate in order to further foster societal acceptance." The tone changes significantly, when turning to a statement by a non-business actor, for instance in the response submitted by the Platform for International Cooperation on Undocumented Migrants (PICUM), an NGO headquartered in Brussels: "We are particularly concerned about the use of AI breaching fundamental rights in the areas of policing and immigration control . . . as well the use of AI in sensitive areas, such as the use of public services without adequate democratic oversight, transparency or evidence to justify the need or purpose of its use." These responses illustrate the reasoning that leads actors to weigh AI concerns differently. Whereas the response by the business actor Thales SA emphasizes that the AI Act should recognize the positive utility of AI, the response by PICUM emphasizes how the application of AI raises important concerns.

We find support also for our second hypothesis that business actors will hold preferences for a less demanding regulatory framework on AI. Figure 3 exhibits the predictions based on our regression models. The predicted level of importance of regulatory stringency for business actors is 4.18, suggesting that this type of actor typically favors a laxer regulatory environment for AI than academic (4.54), NGO (4.72), and other (4.76) actors. Indeed, it is noteworthy that nearly all non-business actors are very close to the maximum value on all dimensions of the regulatory framework considered in the questions included in this analysis. While all types of non-state actors see a need for regulation of AI development that is protective of individual rights, transparent, and incorporates human oversight, business actors are relatively more interested in balancing such protection against room for innovation. These findings support our theoretical intuition that support for innovation versus protection is a question of degree, where business actors and other non-state actors recognize the value of both goals, but strike the balance differently.

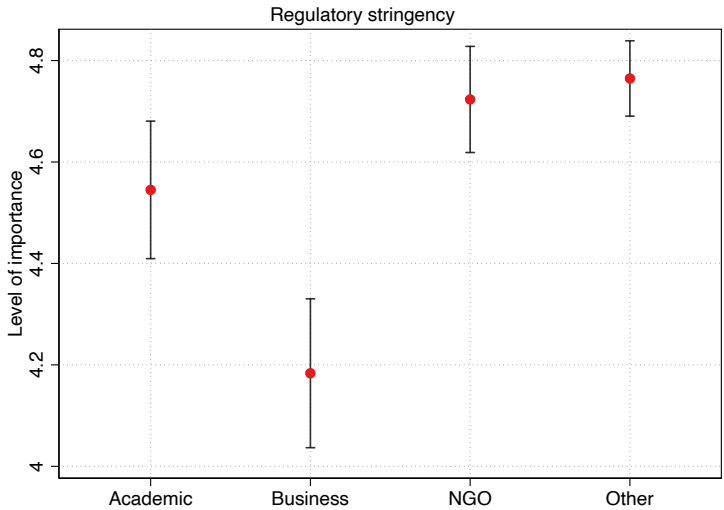

**Figure 3.** Adjusted predictions of group type on regulatory stringency (1-5). Higher values correspond to a preference for more demanding regulation. Average marginal effects with 95 percent confidence intervals. Calculated based on Model 2 in Table A.2. Standard errors clustered on countries. N=427.

The contrast between business actors and other actors are again reflected in the qualitative comments submitted during the consultation procedure. For example, the Computer & Communication Industry Association Europe, a business association, stresses that introducing strict liability for AI "would have a chilling effect on innovation, increase development costs and the uptake of AI," whereas Digital Europe, an organization representing the digital industry, argues that the formulation of the AI Act need to "avoid burdensome requirements for companies serving markets across the world." Conversely, many NGO submissions point to the need to for strong oversight and regulation. For example, PICUM's submission argues that compliance with a prospective AI Act "must be evaluated by a trusted external actor, and not on the basis of self-regulation" whereas the All European Trade Union wants to include provisions to "mandate that any machine learning software taking decisions regarding humans and specifically workers or embedded in a safety-critical system be explainable - and prohibit its use if not the case." In general, business actors tend to favor an AI Act with fewer mandatory requirements and a higher degree of self-regulation, whereas non-business actors prefer a more stringent mandatory requirements and stronger and more centralized compliance monitoring.

In Figures 4 and 5, we disaggregate results across different types of business categories, focusing on comparing preferences submitted by tech actors to those of business actors in other sectors. As visualized in Figure 4, the estimated concern of firms and business associations active in tech is estimated at 0.4 points lower than that of other groups, but this difference is significant only at the $p < 0.10$ level. This potential difference in concern levels might stem from the familiarity and exposure that tech-focused firms and associations have with AI-related technologies, potentially indicating a greater confidence in their ability to navigate and mitigate associated risks. We also note, however, that there is considerable variance in the estimate for tech actors, suggesting that they are more diverse in their views than other business actors.

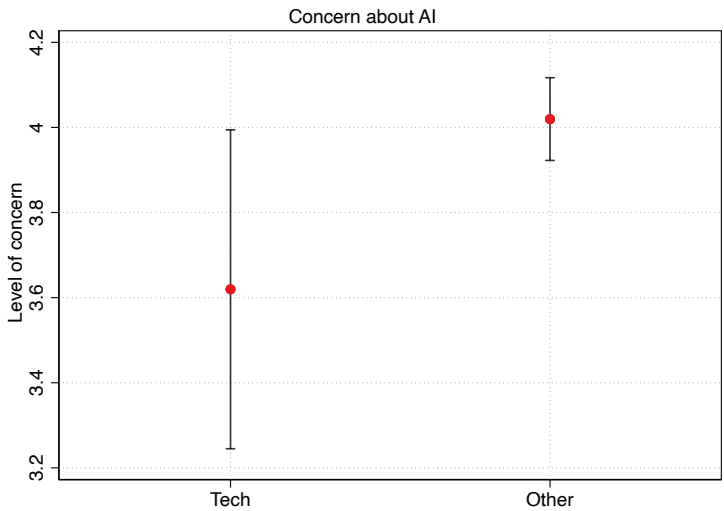

**Figure 4.** Adjusted predictions of level of concern about AI, tech actors compared with other business actors. Average marginal effects with 95 percent confidence intervals. Calculation based on Model 1 in Table A.3. Standard errors clustered on countries. N=158.

While tech actors are less concerned about AI, they are comparable to other business actors when it comes to their preferences for regulatory stringency. As shown in Figure 5, the predicated mean for tech actors (4.06) is comparable to that of non-tech business actors (4.22) and the two estimates are statistically indistinguishable (p=0.4). This finding suggests that despite perceiving lower risks, tech actors advocate as stringent EU legislation as other business groups, possibly to foster a predictable environment within the rapidly evolving AI landscape.

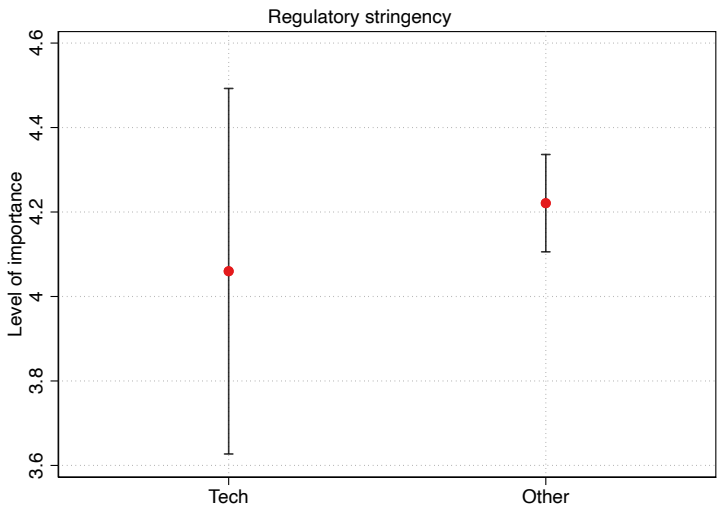

**Figure 5.** Adjusted predictions of regulatory preferences, tech actors compared with other business actors. Average marginal effects with 95 percent confidence intervals. Calculated based on Model 2 in Table A.3. Standard errors clustered on countries. N=168.

Thus far, our analysis has concluded that there are distinct differences between the concerns and regulatory preferences of business actors and other groups that participated in the EU's public consultation on the EU AI Act. We now proceed to investigate whether these differences are conditional on country-level characteristics. Recall that our third and fourth hypotheses were formulated to test the propositions that differences between business actors and non-business actors would be accentuated in countries with more developed AI sectors, both regarding concerns about AI (H3) and regulatory preferences (H4).

Our evidence is supportive of both hypotheses. Figure 6 illustrates that the effect of group type on the level of concern about AI (left) and regulatory preferences (right) varies across the range of the underlying variable, the commercial component of the Global AI index. The substantive effect is non-negligible. Whereas a business actor headquartered in a country with the lowest level of commercial AI development (0) would have a predicted level of concern of about 4.1, an actor based in a country with the highest level (10) would have a predicted value of about 3.8. For regulatory framework, the same shift corresponds to a reduction of predicted values from 4.4 to 4.1. In other words, consistent with our conjecture, there is a tendency to greater dispersion across business and non-business actors as a country's AI industry develops. Submissions from actors located in countries with less developed commercial AI sectors are more similar to each other than actors from countries with more developed sectors.[66]

---

[66] In Figure A6 in the online appendix we predictions that compare tech against non-tech groups.

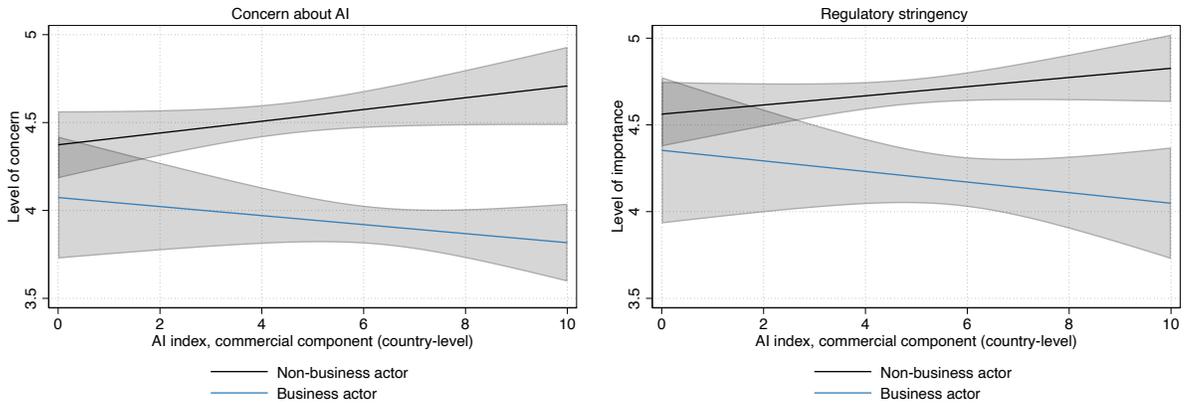

**Figure 6.** Adjusted predictions of group type, conditional on national-level AI index scores. Average marginal effects with 95 percent confidence intervals. Calculated based on Models 3 and 4 in Table A.2. Standard errors clustered on countries. N=419.

*Robustness tests*

In sum, we find that our empirical data are supportive of our theoretical propositions reflected in H1-H4. To ascertain that our results are not driven by particularities of modeling, specification, or data choices, we performed five main types of robustness checks.

First, we evaluated whether our results are an artifact of the creation of the indices for concern about AI and regulatory stringency. We estimated separate regression models for each component of the indices, based on each of the constituent questions in the public consultation questionnaire. Tables A.4. and A.5. present the results, demonstrating that our results are not contingent on including or excluding any particular question. Indeed, there is a high degree of similarity of results across each of the question-specific models.

Second, we used alternative approaches to account for the clustered nature of our data and the possible influence of country-level factors. Tables A.7 in the online appendix present results for a multilevel model with varying intercepts for countries.[67] Table A.8. presents models with country fixed effects, which is another way to account for clustering and country-level confounding, while Table A.9. exhibits results where we explicitly control for country-level observables, including liberal democracy, corruption perceptions, population size, and economic development. The results exhibit no substantive deviation from our main results, and we again observe that business actors deviate significantly from other actors both with regard to concern about AI and regulatory preferences.

Third, while we have theoretical reasons to focus on non-state actors based within the EU, we wanted to ascertain that our results are not driven by the exclusion of non-EU responses. In Table A.10., also in the online appendix, we include responses from non-EU groups. Again, the results are very similar to the EU-only results and we also note (in models 3 and 4) that the difference between business and non-business actors is observed also outside of the EU-based groups.

---

[67] Gelman and Hill (2006).

Fourth, since the views on AI of organizations headquartered in Belgium, where most of the key institutions of the EU are located, may not be fully comparable to those headquartered in other countries, we estimate key models on samples excluding such organizations. As shown in Table A.11 the results are insensitive to excluding groups based in Belgium.

Fifth, we examined whether the results were driven by grouping firms and business associations together in the business actor category. As can be seen in Figure A.5. in the online appendix, we find no such evidence. While firms are somewhat less concerned about AI than business associations, both types are robustly distinct from other non-state actors, and they are comparable in terms of regulatory preferences.

## Conclusion

The EU AI Act will introduce a common European regulatory framework for AI technology. Because of its expected far-reaching consequences, the proposed Act has attracted considerable attention from non-state actors trying to influence the terms and conditions of the new framework. In this article, we have offered the first systematic analysis of non-state actor preferences toward international regulation of AI, focusing on the case of the EU AI Act. Theoretically, we have developed an argument about the varying concerns and preferences of business actors and other non-state actors with respect to AI technology and its regulation. Empirically, we have tested our argument using data from the public consultations organized by the European Commission in 2020, conducting descriptive and regression analyses of the expressed concerns and regulatory preferences of non-state actors.

Our principal results are threefold. First, we find that all types of non-state actors express concerns about AI technology and are in favor of regulating its development and use at the European level. Second, as expected, we nonetheless observe significant variation across types of non-state actors, both with regard to expressed concerns and regulatory preferences. Business actors tend to favor a laxer regulatory environment compared to other non-state actors, privileging innovation over protection. Third, we find that the strength of the commercial AI sector in a country affects the differences between business actors and other types of non-state actors. In countries where the commercial AI sector is more developed, the differences in concerns and preferences between business actors and other non-state actors become more pronounced.

While this article contributes important new evidence on non-state actor preferences toward AI regulation, we should also note the study's limitations and how future research might address them. For one thing, we have worked with a simplified dichotomy between business actors and other non-state actors, while also distinguishing between tech companies and other businesses in the first category. Future research may contribute further fine-grained analyses of the specific preferences of additional sub-types of non-state actors. Furthermore, future research could seek to broaden the scope of the studied non-state actors beyond those that participate actively in public consultations. While participation in a consultation procedure is indicative of an interest to influence AI regulation, we cannot exclude that some non-state actors choose other channels for expressing their concerns and preferences. Finally, future research could assess the generalizability of these findings by conducting similar analyses of non-state actor preferences toward AI regulation in other international settings. The Council of Europe (CoE), the Group of Seven (G7), the Organization of Economic Cooperation and Development (OECD), and the United Nations (UN) are all engaged in developing principles for the development and use of AI technology.

Yet, for now, our findings carry several broader implications for research and policy. First, this article contributes new knowledge on preferences toward AI regulation, complementing previous studies examining the preferences of states[68] and citizens[69]. The article's findings on non-state actor preferences point to important similarities and differences across actor categories: much like citizens, non-state actors in general are concerned about the risks of AI and quite supportive of regulation; and much like states, non-state actors are divided in the relative importance they assign innovation and protection in the regulation of AI.

---

[68] E.g., Radu (2021); Djeffal et al. (2022).
[69] E.g., Zhang (2023); König et al. (2023).

Second, our study adds to the small but swiftly growing field of research on regional and global AI governance[70] (for overviews, see Dafoe 2018; Tallberg et al. 2023). Previous research on AI governance beyond the nation state has tended to focus on the emerging global AI regime[71] (Butcher and Beridze 2019; Schmitt 2021), institutional designs for the governance of AI[72] (Cihon et al. 2020), and key principles guiding AI regulation[73] (Jobin et al. 2019). In contrast, this article privileges non-state actors, showing how such actors demand international regulation of AI, but hold varying preferences about the appropriate balance between business innovation and public protection.

Finally, our results shed light on the types of interest conflicts that policymakers need to confront when developing AI regulation. Non-state actor support is likely critical for AI regulation to be effective and legitimate. Our analysis shows that policymakers need to balance the competing concerns and preferences of business actors, on the one hand, and NGOs, research institutes, and labor unions, on the other hand. In addition, it raises important knock-on questions about the influence of competing non-state actors on state positions in multilateral negotiations and on international regulatory outcomes. As the most comprehensive regulatory framework worldwide, the EU AI Act presents a scientifically valuable and politically important case for exploring these issues.

---

[70] For overviews, see Dafoe (2018); Tallberg et al. (2023).
[71] Butcher and Beridze (2019); Schmitt (2021).
[72] Cihon et al. (2020).
[73] Jobin et al. (2019).

# Appendix

## A.1. Extract from public consultation questionnaires

### Questions pertaining to concerns about AI

| F25 | In your opinion, how important are the following concerns about AI (1-5: 1 is not important at all, 5 is very important)?<br>: AI may endanger safety |
|---|---|
| F26 | In your opinion, how important are the following concerns about AI (1-5: 1 is not important at all, 5 is very important)?<br>: AI may breach fundamental rights (such as human dignity, privacy, data protection, freedom of expression, workers' rights etc.) |
| F27 | In your opinion, how important are the following concerns about AI (1-5: 1 is not important at all, 5 is very important)?<br>: The use of AI may lead to discriminatory outcomes |
| F28 | In your opinion, how important are the following concerns about AI (1-5: 1 is not important at all, 5 is very important)?<br>: AI may take actions for which the rationale cannot be explained |
| F29 | In your opinion, how important are the following concerns about AI (1-5: 1 is not important at all, 5 is very important)?<br>: AI may make it more difficult for persons having suffered harm to obtain compensation |
| F30 | In your opinion, how important are the following concerns about AI (1-5: 1 is not important at all, 5 is very important)?<br>: AI is not always accurate |

### Questions pertaining to regulatory stringency

| F39 | In your opinion, how important are the following mandatory requirements of a possible future regulatory framework for AI (as section 5.D of the White Paper) (1-5: 1 is not important at all, 5 is very important)?: The quality of training data sets |
|---|---|
| F40 | In your opinion, how important are the following mandatory requirements of a possible future regulatory framework for AI (as section 5.D of the White Paper) (1-5: 1 is not important at all, 5 is very important)?: The keeping of records and data |
| F41 | In your opinion, how important are the following mandatory requirements of a possible future regulatory framework for AI (as section 5.D of the White Paper) (1-5: 1 is not |
| | important at all, 5 is very important)?: Information on the purpose and the nature of AI systems |
| F42 | In your opinion, how important are the following mandatory requirements of a possible future regulatory framework for AI (as section 5.D of the White Paper) (1-5: 1 is not important at all, 5 is very important)?: Robustness and accuracy of AI systems |
| F43 | In your opinion, how important are the following mandatory requirements of a possible future regulatory framework for AI (as section 5.D of the White Paper) (1-5: 1 is not important at all, 5 is very important)?: Human oversight |
| F44 | In your opinion, how important are the following mandatory requirements of a possible future regulatory framework for AI (as section 5.D of the White Paper) (1-5: 1 is not important at all, 5 is very important)?: Clear liability and safety rules |

## A.2. Regression estimates, concerns about AI and regulatory stringency

| | (1) | (2) | (3) | (4) |
|---|---|---|---|---|
| | AI concern | Regulatory stringency | AI concern | Regulatory stringency |
| | | | | |
| Business actor | -0.47*** | -0.36*** | -0.30 | -0.21 |
| | (0.09) | (0.09) | (0.19) | (0.21) |
| | | | | |
| NGO | 0.21 | 0.18* | | |
| | (0.12) | (0.08) | | |
| | | | | |
| Other | 0.18 | 0.22** | | |
| | (0.11) | (0.07) | | |
| | | | | |
| AI index, commercial component | | | 0.03 | 0.03 |
| | | | (0.02) | (0.02) |
| | | | | |
| Business actor × AI index, commercial component | | | -0.06* | -0.06 |
| | | | (0.03) | (0.03) |
| | | | | |
| Constant | 4.39*** | 4.55*** | 4.37*** | 4.56*** |
| | (0.09) | (0.06) | (0.09) | (0.09) |
| N | 411 | 427 | 404 | 419 |
| $R^2$ | 0.19 | 0.18 | 0.19 | 0.19 |

Robust errors clustered on countries in parenthesis. Academic non-state actors are reference category in models 1 and 2. Non-business actors reference group in models 3 and 4. Two-tailed tests.* $p < 0.05$, ** $p < 0.01$, *** $p < 0.001$

### A.3. Regression estimates, concerns about AI and regulatory stringency, sample of business actors

| | (1) | (2) | (3) | (4) |
|---|---|---|---|---|
| | AI concern | Regulatory stringency | AI concern | Regulatory stringency |
| | | | | |
| Tech actor | 0.40 | 0.16 | -0.59 | 0.10 |
| | (0.19) | (0.20) | (0.44) | (0.51) |
| | | | | |
| AI index, commercial component | | | -0.17** | -0.05 |
| | | | (0.06) | (0.09) |
| | | | | |
| Tech actor × AI index, commercial component | | | 0.19** | 0.02 |
| | | | (0.06) | (0.08) |
| | | | | |
| Constant | 3.62*** | 4.06*** | 4.55*** | 4.31*** |
| | (0.18) | (0.21) | (0.41) | (0.51) |
| $N$ | 158 | 168 | 156 | 166 |
| $R^2$ | 0.061 | 0.010 | 0.119 | 0.023 |

Robust errors clustered on countries in parenthesis. Non-tech actors are reference category. Two-tailed tests. * $p < 0.05$, ** $p < 0.01$, *** $p < 0.001$

## A.4. Regression estimates, individual questionnaire components relating to AI concern

| | (1) | (2) | (3) | (4) | (5) | (6) |
|---|---|---|---|---|---|---|
| Question | F25 | F26 | F27 | F28 | F29 | F30 |
| | | | | | | |
| Business actor | -0.24[*] | -0.33[**] | -0.33[**] | -0.60[***] | -0.56[***] | -0.58[***] |
| | (0.09) | (0.09) | (0.10) | (0.15) | (0.14) | (0.12) |
| | | | | | | |
| NGO | 0.38[*] | 0.18 | 0.18 | 0.04 | 0.41[*] | 0.14 |
| | (0.14) | (0.12) | (0.11) | (0.17) | (0.16) | (0.16) |
| | | | | | | |
| Other | 0.35[***] | 0.07 | 0.11 | 0.01 | 0.31[*] | 0.31[*] |
| | (0.08) | (0.13) | (0.10) | (0.11) | (0.14) | (0.11) |
| | | | | | | |
| Constant | 4.26[***] | 4.63[***] | 4.57[***] | 4.44[***] | 4.11[***] | 4.15[***] |
| | (0.11) | (0.10) | (0.09) | (0.12) | (0.12) | (0.11) |
| N | 446 | 451 | 446 | 448 | 433 | 440 |
| $R^2$ | 0.103 | 0.081 | 0.080 | 0.110 | 0.180 | 0.129 |

Robust errors clustered on countries in parenthesis. Academic non-state actors are reference category. See Table A1 for explanation of F25-F30. Two-tailed tests. $^{*}$ $p < 0.05$, $^{**}$ $p < 0.01$, $^{***}$ $p < 0.001$.

## A.5. Regression estimates, individual questionnaire components relating to stringency of regulation

| | (1) | (2) | (3) | (4) | (5) | (6) |
|---|---|---|---|---|---|---|
| Question | F39 | F40 | F41 | F42 | F43 | F44 |
| | | | | | | |
| Business actor | -0.46*** | -0.31* | -0.23* | -0.27* | -0.55*** | -0.31** |
| | (0.10) | (0.14) | (0.10) | (0.13) | (0.11) | (0.10) |
| | | | | | | |
| NGO | 0.15 | 0.16 | 0.27** | 0.14 | 0.20* | 0.17 |
| | (0.12) | (0.12) | (0.08) | (0.14) | (0.07) | (0.09) |
| | | | | | | |
| Other | 0.21 | 0.19 | 0.21* | 0.35* | 0.15 | 0.25** |
| | (0.13) | (0.10) | (0.09) | (0.13) | (0.09) | (0.07) |
| | | | | | | |
| Constant | 4.58*** | 4.42*** | 4.49*** | 4.52*** | 4.60*** | 4.67*** |
| | (0.10) | (0.12) | (0.06) | (0.10) | (0.06) | (0.07) |
| N | 443 | 445 | 448 | 447 | 448 | 445 |
| $R^2$ | 0.119 | 0.070 | 0.082 | 0.101 | 0.162 | 0.129 |

Robust errors clustered on countries in parenthesis. Academic non-state actors are reference category. See Table A1 explanation of F39-F44. Two-tailed tests. * $p < 0.05$, ** $p < 0.01$, *** $p < 0.001$.